\documentclass[twocolumn,aps,pre,longbibliography%
]{revtex4-2}
\usepackage{psfrag,epsfig,amsfonts,amssymb,amsmath,wasysym,bm}
\usepackage{dcolumn}
\usepackage{bbold}
\usepackage[normalem]{ulem}
\usepackage{xcolor}
\usepackage{tabularx, tabulary, booktabs}
\usepackage{longtable}
\usepackage{tabu}
\usepackage{enumitem}
\usepackage{hyperref}

\newcolumntype{Z}{>{\raggedleft\arraybackslash}X}

\newcommand{\<}{\langle}	
\renewcommand{\>}{\rangle}	

\newcommand{\var}{\operatorname{var}}
\newcommand{\cov}{\operatorname{cov}}

\newcommand{\e}{\mathrm{e}}		
\newcommand{\I}{\mathrm{i}}		

\renewcommand{\d}{\mathrm{d}} 
\newcommand{\lmat}{\left( \begin{matrix}}	
\newcommand{\rmat}{\end{matrix} \right)}	
\newcommand{\tr}{\operatorname{tr}} 

\newcommand{\RR}{{\mathbb R}}

\newcommand{\NN}{{\mathbb N}}

\newcommand{\Part}{\mathcal{P}}

\begin{document}

\title{Temperature and conditions for thermalization after canonical quenches}
\author{Lennart Dabelow}
\email{l.dabelow@qmul.ac.uk}
\affiliation{School of Mathematical Sciences, Queen Mary University of London, London E1~4NS, UK}

\date{\today}

\begin{abstract}
We consider quenches of a quantum system that is prepared in a canonical equilibrium state of one Hamiltonian
and then evolves unitarily in time under a different Hamiltonian.
Technically, our main result is a systematic expansion of the pre- and post-quench canonical ensembles in the quench strength.
We first demonstrate how this can be used to predict the system's temperature after the quench from equilibrium properties at the pre-quench temperature.
For a thermalizing post-quench system,
it furthermore allows us to calculate equilibrium observable expectation values.
Finally,
in the presence of additional conserved quantities besides the Hamiltonian, 
we obtain a hierarchy of necessary conditions for thermalization towards the (post-quench) canonical ensemble.
At first order,
these thermalization conditions have a nice geometric interpretation in operator space with the canonical covariance as a semi-inner product:
The quench operator (difference between post- and pre-quench Hamiltonians) and the conserved quantity must be orthogonal
in the orthogonal complement of the post-quench Hamiltonian.
We illustrate the results numerically for a variety of setups involving integrable and nonintegrable models.
\end{abstract}

\maketitle

\section{Introduction}
\label{sec:Introduction}

If left alone,
generic systems of our everyday experience tend towards thermal equilibrium,
a state that is characterized by macroscopically homogeneous and stationary observable properties.
Through thermodynamic ensembles, these equilibrium properties can be predicted from a few macroscopic parameters, such as the temperature, number of particles, and volume of the system.
Understanding this macroscopically ubiquitous phenomenon based on the fundamental laws of quantum mechanics that describe the system's microscopic constituents
has been of significant research interest in recent years \cite{Gogolin2016ete, DAlessio2016qce, Mori2018tpi, Ueda2020qet}.

A common way to prepare such a system out of equilibrium
is a so-called quench:
We start with a system in equilibrium and perform a fast, virtually instantaneous change of some of its parameters,
such as suddenly moving a piston or switching an external field.
This manipulation typically changes the system's energy and temperature,
but it is not straightforward to predict the latter, in particular.

After the quench,
the system usually relaxes towards a new (quasi)stationary state.
Generically, expectation values of physical observables at sufficiently late times
can be characterized by one of the standard thermodynamic ensembles (microcanonical, canonical, ...)
and 
we say that the system \emph{thermalizes}.
However, situations where this paradigm is seemingly violated
and the long-time behavior retains some memory of the initial conditions \cite{Yurovsky2011mic, Gogolin2011atn, Bartsch2017net}
have been of considerable interest recently, too.
Mechanisms that can result in absence of thermalization
include integrability \cite{Essler2016qdr, Vidmar2016gge},
many-body localization \cite{Nandkishore2015mbl, Abanin2019cmb, Sierant2025mbl},
many-body scars \cite{Serbyn:2021qmb, Moudgalya2022qmb, Chandran2023qmb},
and Hilbert space fragmentation \cite{Moudgalya2022qmb, Chandran2023qmb}.
It is widely agreed that conserved quantities,
inherent or emergent,
play a major role in understading thermalization or its absence \cite{Essler2016qdr, Vidmar2016gge, Nandkishore2015mbl, Abanin2019cmb, Sierant2025mbl, Serbyn:2021qmb, Moudgalya2022qmb, Chandran2023qmb, Majidy2023ncc}.
In other words, they need to be taken into account properly
to characterize the system's equilibrium properties correctly.
Then again,
the number of independent conserved quantities of any quantum system equals its Hilbert space dimension (take, e.g., the projectors onto energy eigenstates),
which is exponentially large in the system's degrees of freedom.
However, there are no known, physically realistic setups in which such an exponentially large number of conservation laws is necessary to predict the long-term behavior of a many-body system.
This raises the question of which conserved quantities are relevant,
meaning that they need to be considered to describe the system's equilibrium properties,
and which are not.
There are a variety of qualitative and quantitative criteria that assess the importance of conserved quantities for thermalization,
such as locality \cite{Fagotti2013rdm, Pozsgay:2017gge}, overlap with observables of interest \cite{Mierzejewski2020qii}, Mazur bounds \cite{Mazur1969nep, Mierzejewski2014bgg, Dhar2021rmb, Moudgalya2024ecq}, the quench action \cite{Caux2016qa}, quasiparticle excitations \cite{Ilievski2017ipe}, and commutant algebras \cite{Moudgalya2022hsf, Moudgalya2024ecq},
for example.
Nevertheless, a comprehensive understanding of this issue is still missing
(and, to avoid misunderstandings, will not be provided here, either).
A related problem is to find potentially relevant and physically meaningful (e.g., local or quasilocal) conserved quantities \cite{Ilievski2015qco, Ilievski2015cgg, Lydzba2024lim, Zhan2024lcl, Shtanko2025uli}.

The present work contributes to the ongoing efforts to understand thermalization and the role of conserved quantities during this process.
We focus on quenches from canonical equilibrium states
and address the following three questions, in particular:
(i) Assuming that the system thermalizes,
what are its temperature and equilibrium properties after the quench?
(ii) If the post-quench system has additional conserved quantities besides the Hamiltonian,
what are necessary conditions for thermalization towards the canonical ensemble?
(iii) Under what circumstances must a conserved quantity be accounted for explicitly (e.g., through a 
generalized Gibbs ensemble \cite{Rigol:2007rci, Langen2015eog, Vidmar2016gge})
to reproduce the system's equilibrium properties?
Our (partial) answers to these questions will be based on a systematic expansion of post-quench properties like the temperature and observable expectation values in the quench strength.

\section{Setup}
\label{sec:Setup}

We focus on isolated quantum systems that evolve unitarily in time.
Before the quench,
the system of interest is described by the pre-quench Hamiltonian $\tilde H$
and finds itself in thermal equilibrium,
which we model by an initial state in the form of the canonical ensemble
(or Gibbs state) \cite{note:CanonicalEnsemble} 
at the pre-quench inverse temperature $\tilde\beta$,
\begin{equation}
\label{eq:rho0}
	\tilde\rho_{\tilde\beta} := \tilde Z_{\tilde\beta}^{-1} \, \e^{-\tilde\beta \tilde H}
	\quad\text{with}\quad
	\tilde Z_{\tilde\beta} := \tr( \e^{-\tilde\beta \tilde H} ) \,.
\end{equation}
Note that, thanks to dynamical typicality \cite{Bartsch2009dtq, Sugiura2013ctp, Reimann2020wam},
all of the following conclusions equally apply to large classes of pure initial states with equilibrium-like properties.
Furthermore, other thermodynamic ensembles (e.g., microcanonical) are also expected to result in similar conclusions, at least for systems with short-range interactions, as ensembles are known to be equivalent under rather general conditions \cite{Touchette2015ene, Brandao2015esm, Tasaki:2018lec, Kuwahara2020etc, Kuwahara2020gcb},
though exceptions exists (see, e.g., Refs.~\cite{Campa2009smd, Russomanno2021qce, Defenu2024eil, Vardi2024nts}).

After the quench,
the system is described by the post-quench Hamiltonian
\begin{equation}
	H := \tilde H + g V \,.
\end{equation}
We call $V$ the \emph{quench operator} and $g$ the (dimensionless) \emph{quench strength}.
In the absence of further (relevant) conserved quantities besides $H$,
a generic isolated system is expected to thermalize,
meaning that
its long-term properties are characterized by the canonical ensemble of $H$,
\begin{equation}
\label{eq:rho:PostQuench}
	\rho_{\beta} := Z_{\beta}^{-1} \, \e^{-\beta H}
	\quad\text{with}\quad
	Z_{\beta} := \tr( \e^{-\beta H} ) \,,
\end{equation}
where the post-quench inverse temperature $\beta$ is fixed by the condition
\begin{equation}
\label{eq:PostQuenchTemp:Condition}
	\< H \>_{\! \tilde\rho_{\tilde\beta}} = \< H \>_{\!\rho_\beta} \,.
\end{equation}
Here $\< A \>_{\!\rho} := \tr(\rho A)$ denotes the expectation value of the observable $A$ in the state $\rho$.

The question of whether and when the system indeed thermalizes towards such a canonical state provides the overarching context of the present study.
However, we will not address the related question of equilibration,
i.e., whether the long-term behavior becomes macroscopically stationary,
such that physically realistic observables do not change noticeably over time (apart from technically unavoidable, but extremely rare quantum revival effects).
This can be shown to happen very generally under relatively weak assumptions \cite{Short2012qef, Reimann2012eim, Balz2016eim, Chiba2023kol}
and will be tacitly taken for granted in the following.
Likewise, we will not be concerned with time scales or any other dynamical properties of the relaxation process.

Methodologically, the main idea is to expand the post-quench inverse temperature $\beta$ and equilibrium expectation values $\< A \>_{\rho_\beta}$ of an observable $A$ in powers of $g$.

\section{Post-quench temperature and expectation values}
\label{sec:PostQuenchTemperature}

Our first main result is a perturbative calculation of the system temperature and further equilibrium properties after the quench.
To this end, we solve Eq.~\eqref{eq:PostQuenchTemp:Condition} for the post-quench inverse temperature $\beta$.
The goal is to predict $\beta$ for arbitrary quench strengths $g$ from properties of the post-quench system at the known pre-quench temperature $\tilde\beta$.

Our starting point is a formal expansion of $\beta$ in the quench strength $g$,
\begin{equation}
\label{eq:PostQuenchTemp:Expansion}
	\beta = \sum_{n=0}^\infty \beta_n g^n \,.
\end{equation}
At zeroth order ($g = 0$, no quench),
we evidently have
\begin{equation}
\label{eq:PostQuenchTemp:0}
	\beta_0 = \tilde\beta
\end{equation}
because $H = \tilde{H}$ in~\eqref{eq:PostQuenchTemp:Condition},
i.e., the zeroth order post-quench temperature is the pre-quench temperature.
The corrections $\beta_n$ for $n \geq 1$ can be obtained recursively in two steps.

First, we express expectation values $\< A \>_{\!\rho_\beta}$ of an arbitrary observable $A$ in the post-quench ensemble at the unknown temperature $\beta$ in terms of canonical expectation values at the known pre-quench temperature $\tilde\beta = \beta_0$.
To this end,
we formally write 
$\e^{-(\beta - \beta_0) H}$
as a power series in $g$ using a multinomial expansion.
This result can then be exploited, on the one hand, to calculate $\tr[\e^{-\beta H} A]$ as a series in $g$ with coefficients involving factors of $\tr[\e^{-\beta_0 H} H^k A]$.
On the other hand,
it can be used to represent the ratio of partition functions $Z_{\beta_0} / Z_{\beta}$ as a geometric series in $g$.
Multiplying these two expressions and sorting by powers of $g$,
we obtain the desired expansion of $\< A \>_{\!\rho_\beta}$ in terms of combinations of $\< H^k A \>_{\!\rho_{\beta_0}}$.
To order $g^2$, it reads
\begin{equation}
\label{eq:PostQuench:Expv:Leading}
\begin{aligned}
	& \< A \>_{\!\rho_\beta}
		= \< A \> - g \beta_1 \cov(H, A) \\
		&\quad	- g^2 \left[ (\beta_2 + \beta_1^2 \<H\>) \cov(H, A) - \tfrac{\beta_1^2}{2} \cov(H^2, A) \right] 
			+ \ldots \,,
\end{aligned}
\end{equation}
where
\begin{equation}
	\< A \> := \< A \>_{\!\rho_{\beta_0}} = Z_{\beta_0}^{-1} \tr(\e^{-\beta_0 H} A)
\end{equation}
denotes expectation values of the post-quench ensemble at the pre-quench reference temperature $\beta_0$,
and
\begin{equation}
	\cov(A, B) := \< A B \> - \< A \> \< B \>
\end{equation}
is the corresponding covariance.
In other words, we suppress the canonical ensemble in the notation for expectation values and covariances if it corresponds to the post-quench system at the pre-quench temperature;
in all other cases, the state will still be indicated explicitly as a subscript of the expectation-value brackets.
Eq.~\eqref{eq:PostQuench:Expv:Leading} and the full expansion beyond second order are derived in detail in Appendix~\ref{app:DifferentTemperatures},
see Eq.~\eqref{eq:PostQuench:Expv2} for the final result.

Second, we express expectation values $\< A \>_{\!\tilde\rho_{\tilde\beta}}$ in the pre-quench ensemble in terms of post-quench ensemble expectation values and correlation functions.
In this case, we expand $\e^{-\tilde\beta \tilde H} = \e^{-\tilde\beta(H - g V)}$ in $g$,
resulting in an imaginary-time Dyson series.
Similarly as in the previous step,
we can adopt this result to calculate series expansions of $\tr[\e^{-\tilde \beta \tilde H} A]$ and $Z_{\tilde\beta} / \tilde Z_{\tilde\beta}$,
combine the expressions and sort by powers of $g$
to find the expansion of $\< A \>_{\!\tilde\rho_{\tilde\beta}}$.
To order $g^2$, we obtain
\begin{equation}
\label{eq:PreQuench:Expv:Leading}
\begin{aligned}
	\< A \>_{\!\tilde\rho_{\tilde\beta}}
		&= \< A \> + g \cov(\Phi_1, A) \\
		&\quad
			+ g^2 \left[ \cov(\Phi_2, A) - \< \Phi_1 \> \cov(\Phi_1, A) \right]
			+ \ldots \,,
\end{aligned}
\end{equation}
where
\begin{equation}
	\Phi_n := \int_0^{\tilde\beta} \!\! \d\lambda_1 \int_0^{\lambda_1} \!\! \d\lambda_2 \cdots \int_0^{\lambda_{n-1}} \!\! \d\lambda_n \, V(\lambda_1) \cdots V(\lambda_n)
\end{equation}
is the $n$th-order coefficient of the time-ordered exponential of the imaginary-time-evolved quench operator $V(\lambda) := \e^{\lambda H} V \e^{-\lambda H}$.
The derivation of Eq.~\eqref{eq:PreQuench:Expv:Leading} can be found in Appendix~\ref{app:DifferentHamiltonians}
with the full expansion given in Eq.~\eqref{eq:PreQuench:Expv}.
It should be noted that Eq.~\eqref{eq:PreQuench:Expv:Leading}
is effectively also a high-temperature (small-$\tilde\beta$) expansion because of the expanded exponential.
	
To extract the coefficients $\beta_n$ of the temperature series~\eqref{eq:PostQuenchTemp:Expansion},
we substitute $A = H$
and,
according to Eq.~\eqref{eq:PostQuenchTemp:Condition},
compare Eqs.~\eqref{eq:PostQuench:Expv:Leading} and~\eqref{eq:PreQuench:Expv:Leading} order by order in $g$.
For the first-order correction,
we obtain
\begin{equation}
\label{eq:PostQuenchTemp:1}
	\beta_1 = -\beta_0 \frac{\cov(H, V)}{\var(H)} \,,
\end{equation}
exploiting that $\cov(\Phi_1, H) = \beta_0 \cov(H, V)$.
Here $\var(H) := \cov(H, H)$ denotes the canonical ensemble variance at the pre-quench temperature.
Eq.~\eqref{eq:PostQuenchTemp:1} coincides with the post-quench temperature correction derived in Ref.~\cite{Chiba2023kol} via linear response theory.

At second order, we find
\begin{equation}
\label{eq:PostQuenchTemp:2}
	\beta_2 = \frac{ \frac{\beta_1^2}{2} \cov(H^2, H) - \cov(H, \Phi_2) }{ \var(H)}
		- \beta_1 (\beta_1 \< H \> + \beta_0 \< V \>) \,.
\end{equation}
Note that $\cov(H, \Phi_2)$ can be simplified slightly by writing it as a single integral,
\begin{equation}
\label{eq:CovHPhi2}
	\cov(H, \Phi_2) = \int_0^{\beta_0} \!\! \d\lambda \, \lambda \< V(\lambda) [ H - \< H \> ] V \> \,.
\end{equation}
Higher-order corrections $\beta_n$ can be calculated recursively in principle,
see Appendix~\ref{app:PostQuenchTemperature} and particularly Eq.~\eqref{eq:PostQuenchTemp:n} for an explicit formula.
Due to the high-temperature character of the expansion~\eqref{eq:PreQuench:Expv:Leading},
the temperature prediction via Eqs.~\eqref{eq:PostQuenchTemp:1}, \eqref{eq:PostQuenchTemp:2}, and~\eqref{eq:PostQuenchTemp:n} is expected to work better for smaller values of $\tilde\beta = \beta_0$, too.

In summary, Eqs.~\eqref{eq:PostQuenchTemp:1}, \eqref{eq:PostQuenchTemp:2}, and~\eqref{eq:PostQuenchTemp:n} describe a systematic way to estimate the temperature after a canonical quench from equilibrium properties of the system at the pre-quench temperature,
with increasing precision as higher orders of the expansion are taken into account.
Provided that the system thermalizes after the quench,
the combination of Eq.~\eqref{eq:PostQuench:Expv:Leading} [see also Eq.~\eqref{eq:PostQuench:Expv2}]
with Eqs.~\eqref{eq:PostQuenchTemp:0}, \eqref{eq:PostQuenchTemp:1}, \eqref{eq:PostQuenchTemp:2} [and~\eqref{eq:PostQuenchTemp:n} for higher orders] additionally offers a systematic way to calculate post-quench equilibrium expectation values of arbitrary observables $A$
order by order in the quench strength.
In other words, it allows us to predict general equilibrium properties after the quench from equilibrium properties at the pre-quench temperature.

\section{Necessary conditions for thermalization}
\label{sec:ThermalizationConditions}

Next, we assume that there is a physical observable $Q$ (e.g., a sum of local and few-body operators) that commutes with the post-quench Hamiltonian, $[H, Q] = 0$,
so $Q$ is an additional local conserved quantity of the post-quench system besides $H$ itself.
Since $Q$ is conserved,
the system's long-term properties will be described correctly by the post-quench canonical ensemble~\eqref{eq:rho:PostQuench} only if
\begin{equation}
\label{eq:PostQuenchQ:Condition}
	\< Q \>_{\!\tilde\rho_{\tilde\beta}} = \< Q \>_{\!\rho_\beta} \,,
\end{equation}
i.e., only if the initial expectation value of $Q$ coincides with its canonical expectation value.
Note that this is trivially fulfilled if $\tilde\beta = 0$ (and thus $\beta = 0$).
We therefore assume $\tilde\beta > 0$ from now on;
similar conclusions can be drawn for $\tilde\beta < 0$.

Our second main result is a sequence of conditions on the post-quench expectation values and correlation functions that are necessary for the system to thermalize towards the canonical ensemble~\eqref{eq:rho:PostQuench}:
Using the perturbative expansions~\eqref{eq:PostQuench:Expv:Leading} and~\eqref{eq:PreQuench:Expv:Leading},
we conclude that Eq.~\eqref{eq:PostQuenchQ:Condition} must be satisfied at each order in the quench strength $g$.
Consequently,
if any of the resulting conditions is violated,
then the system will fail to thermalize
since $Q$ itself is a trivial example of a nonthermalizing, physical observable.

From a different viewpoint,
we may conclude that the conserved quantity $Q$ is \emph{relevant}
in a given setup
and needs to be considered to characterize the system's equilibrium properties.
Instead of the canonical ensemble~\eqref{eq:rho:PostQuench},
one thus needs to adopt a
generalized Gibbs ensemble
that explicitly accounts for $Q$.

At first order,
the necessary condition for thermalization obtained this way is
\begin{equation}
\label{eq:CondTherm:1}
	\cov(Q, V) \var(H) = \cov(H, Q) \cov(H, V) \,.
\end{equation}

\begin{figure}
	\includegraphics[scale=0.65]{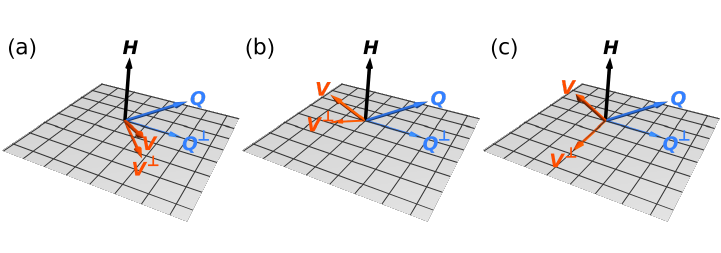}
	\caption{Geometric interpretation of the first-order thermalization condition~\eqref{eq:CondTherm:1:Geometric} [see also Eq.~\eqref{eq:CondTherm:1}].
		(a) No special relationship between $H$, $Q$, and $V$;
		the condition is violated.
		(b) $V$ is orthogonal to $Q$ (with respect to $\cov(\cdot, \cdot)$),
		but the respective projections $V^\perp$ and $Q^\perp$ onto the orthogonal complement of $H$ are not orthogonal to each other;
		the condition is violated.
		(c) The projections $V^\perp$ and $Q^\perp$ are orthogonal (and $V$ and $Q$ may or may not be orthogonal to each other);
		the condition is satisfied.}
	\label{fig:LIOMCond:1:Geometric}
\end{figure}

This condition has an interesting geometric interpretation.
The canonical covariance $\cov(A, B) = \tr(\rho_{\beta_0} A B) - \tr(\rho_{\beta_0} A) \tr(\rho_{\beta_0} B)$ defines an inner-product-like structure on the space of self-adjoint operators
as it is linear in both arguments, positive semidefinite,
and positive definite modulo the real numbers;
cf.\ Appendix~\ref{app:CovarianceInnerProduct} for details.
However, it is generally not symmetric in its arguments,
even though it is for all terms occurring in Eq.~\eqref{eq:CondTherm:1} since both $H$ and $Q$ commute with $\rho_{\beta_0}$.

For an arbitrary observable $A$,
we can define its projection onto the orthogonal complement of the post-quench Hamiltonian $H$ as
\begin{equation}
	A^\perp := A - \frac{ \cov(H, A) }{ \var(H) } H \,.
\end{equation}
The first-order thermalization condition can then be rewritten as
\begin{equation}
\label{eq:CondTherm:1:Geometric}
	\cov(Q^\perp, V^\perp) = 0 \,.
\end{equation}
To observe thermalization to the canonical post-quench ensemble,
it is thus necessary that the quench operator be orthogonal to additional conserved quantities
in the orthogonal complement of the post-quench Hamiltonian.
This is illustrated in Fig.~\ref{fig:LIOMCond:1:Geometric}.

At second order,
the condition~\eqref{eq:PostQuenchQ:Condition} requires
\begin{equation}
\label{eq:CondTherm:2}
\begin{aligned}
	& \cov(Q, \Phi_2) - \beta_0^2 \< V \> \cov(Q, V) \\
		&\;= \frac{\beta_1^2}{2} \cov(H^2, Q) - (\beta_2+ \beta_1^2 \< H \>) \cov(H, Q) \,.
\end{aligned}
\end{equation}
In addition to time-independent canonical expectation values,
this condition also incorporates information from imaginary-time correlation functions
via $\cov(Q, \Phi_2) = \int_0^{\beta_0} \d\lambda \, \lambda \<V(\lambda) [Q - \< Q \>] V\>$.

Higher-order conditions can be obtained from the full series expansions of $\< Q \>_{\!\rho_\beta}$ and $\< Q \>_{\!\tilde\rho_{\tilde\beta}}$ given in Eqs.~\eqref{eq:PostQuench:Expv2} and~\eqref{eq:PreQuench:Expv} in the Appendices, respectively.

It should be noted that,
strictly speaking,
the conditions~\eqref{eq:CondTherm:1} and~\eqref{eq:CondTherm:2}, etc.,
can only be expected to hold exactly in the thermodynamic limit.
A finite system will generally exhibit deviations
in the same way that deviations between time-averaged and thermal expectation values of general physical observables are expected.
However,
compared to the natural scale of fluctuations [e.g., $\var(Q)$],
these deviations should become negligibly small as the system size is increased.

Coming back to the question of what constitutes a relevant conserved quantity,
we inspect the special case $V = Q$,
i.e., a quench in the conserved quantity itself.
In this case, the first-order condition~\eqref{eq:CondTherm:1:Geometric} becomes
\begin{equation}
\label{eq:CondTherm:1:Q=V}
	\var(Q^\perp) = 0 \,.
\end{equation}
Assuming that this holds and recalling the geometric interpretation of the canonical covariance as (essentially) an inner product,
Eq.~\eqref{eq:CondTherm:1:Q=V} means that the ``norm'' of $Q^\perp$ vanishes
and thus $H$ and $Q$ are colinear.
More precisely (see also Appendix~\ref{app:CovarianceInnerProduct}),
we can conclude that $H$ and $Q$ agree up to rescaling and shifting by a constant.
This contradicts the (implicit) assumption that $Q$ is an independent conserved quantity.
Put differently,
the first-order condition is violated for quenches in a conserved quantity.

The physical upshot is
that any conserved quantity that can act as a quench operator is \emph{relevant}:
In a scenario where the quench operator coincides (or overlaps) with the conserved quantity,
the latter
must be taken into account to predict the system's equilibrium properties after the quench.
In principle,
this can be done by restricting the Hilbert space to states within a symmetry sector or a macroscopically small window of $Q$ values compatible with the initial condition (microcanonical approach),
or by including a Lagrange parameter (generalized inverse temperature) conjugate to $Q$ (generalized Gibbs ensemble/grandcanonical approach).

Nevertheless, it may still happen that such a generally relevant conserved quantity is irrelevant in a particular setup (i.e., a particular pair of Hamiltonian $H$ and quench operator $V$)
because its expectation value happens to be unchanged by the quench
such that Eq.~\eqref{eq:PostQuenchQ:Condition} holds to all orders in $g$.

\section{Examples}
\label{sec:Examples}

\begin{table}
\caption{Geometric relations between the operators $H$, $V$, and $Q$ in the examples from Sec.~\ref{sec:Examples}.
A tick ($\checkmark$) in an ``$A \!\perp\! B$'' column indicates that $\cov(A, B) = 0$ for all $\beta_0$ [cf.\ below Eq.\eqref{eq:PostQuench:Expv:Leading}];
$Q^\perp \perp V^\perp$ is the first-order thermalization condition~\eqref{eq:CondTherm:1}, see also Fig.~\ref{fig:LIOMCond:1:Geometric};
$\< Q \>_{\!\tilde\rho_{\tilde\beta}} \!\!=\! \< Q \>_{\!\rho_\beta}$ is the full condition~\eqref{eq:PostQuenchQ:Condition}.}
\label{tab:Examples}
\begin{tabularx}{\columnwidth}{c c c c c c c l}
\toprule
$H$ & $V$ & $Q$ & $H \!\perp\! V$ & $H \!\perp\! Q$ & $Q^\perp \!\!\perp\! V^\perp$ & $\< Q \>_{\!\tilde\rho_{\tilde\beta}} \!\!=\! \< Q \>_{\!\rho_\beta}$ & plot \\
\midrule
$H_1$ & $M^z$ & $Q_n^+$ & $-$ & $-$ & $-$ & $-$ & Fig.~\ref{fig:TFIM1} \\
 & & $Q_n^-$ & $-$ & $\checkmark$ & $\checkmark$ & $\checkmark$ & $\;\;-$ \\
 & $J^z$ & $Q_n^+$ & $\checkmark$ & $-$ & $\checkmark$ & $-$ & Fig.~\ref{fig:TFIM2} \\
 &  & $Q_n^-$ & $\checkmark$ & $\checkmark$ & $-$ & $-$ & Fig.~\ref{fig:TFIM2} \\
$H_2$ & $W$ & $M^z$ & $-$ & $\checkmark$ & $-$ & $-$ & Fig.~\ref{fig:XXZ+9S} \\
 & $M^x$ & $M^z$ & $\checkmark$ & $\checkmark$ & $\checkmark$ & $\checkmark$ & Fig.~\ref{fig:XXZ+5S}a \\
$H_3$ & $M^x$ & $M^z$ & $\checkmark$ & $-$ & $\checkmark$ & $-$ & Fig.~\ref{fig:XXZ+5S}b-d \\
\bottomrule
\end{tabularx}
\end{table}

We numerically explore the temperature expansion~\eqref{eq:PostQuenchTemp:Expansion} and the expansions of expectation values~\eqref{eq:PostQuench:Expv:Leading} and~\eqref{eq:PreQuench:Expv:Leading} in a variety of setups.
In particular, we focus on the thermalization conditions~\eqref{eq:CondTherm:1} and~\eqref{eq:CondTherm:2} for conserved quantities
and various ``geometric'' configurations of the operators $H$, $V$, and $Q$
with respect to the canonical-ensemble covariance
as summarized in Tab.~\ref{tab:Examples}.

\begin{figure}
	\includegraphics[scale=1]{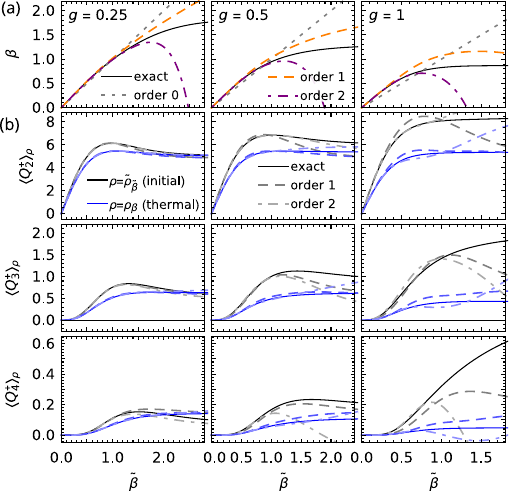}
	\caption{(a) Post-quench inverse temperature $\beta$ and (b) LIOM expectation values $\<Q^+_n\>_{\!\rho}$ as a function of the pre-quench inverse temperature $\tilde\beta$
		for canonical quenches in the transverse-field Ising model~\eqref{eq:H:TFIM}.
		Post-quench: $h = \frac{1}{2}$;
		pre-quench: $h = \frac{1}{2} + g$, $g = 0.25$ (left), $0.5$ (middle), $1$ (right).
		Solid lines are exact values;
		dotted, dashed, and dot-dashed lines correspond, respectively, to the zeroth-, first-, and second-order approximations in the quench strength $g$.
		In (b), gray-shaded lines (with different dashings to indicate different approximation orders, see above) correspond to the initial value $\< Q^+_n \>_{\!\tilde\rho_{\tilde\beta}}$,
		while blue-shaded lines (with different dashings) correspond to the (thermal) expectation value $\< Q^+_n \>_{\!\rho_\beta}$ of the post-quench ensemble,
		where $\beta = \beta(\tilde\beta)$ is the respective post-quench temperature [as shown in panel (a)]. Note that the different line dashings may be hard to distinguish in the regimes where they overlap, indicating good agreement of approximations and exact values.}
	\label{fig:TFIM1}
\end{figure}

\subsection{Ising model}

We begin with two setups where the post-quench Hamiltonian $H = H_1$ is the transverse-field Ising model,
\begin{equation}
	\label{eq:H:TFIM}
	H_1 := - \sum_{j=1}^L \left( \sigma_j^x \sigma_{j+1}^x + h \, \sigma_j^z \right)
\end{equation}
with periodic boundary conditions and $h = \frac{1}{2}$.
Here $\sigma^\nu_j$ ($\nu = x, y, z$) denotes a Pauli operator acting on lattice site $j$.
This model is integrable and exhibits two families of local integrals of motion (LIOM)\cite{Fagotti2013rdm, Essler2016qdr, DAlessio2016qce},
\begin{align}
\label{eq:Q:TFIM:+}
	Q_n^+ &:= - \sum_j \left[ \tau^{xx}_{j,n} + \tau^{yy}_{j,n-2} - h \left( \tau^{xx}_{j,n-1} + \tau^{yy}_{j,n-1} \right) \right] , \\
\label{eq:Q:TFIM:-}
	Q_n^- &:= - \sum_j \left( \tau^{xy}_{j,n} - \tau^{yx}_{j,n} \right) ,
\end{align}
where
$\tau^{\mu\nu}_{j,n} := \sigma^\mu_j \left[ \prod_{k=1}^{n-1} \sigma^z_{j+k} \right] \sigma^{\nu}_{j+n}$ for $n \geq 1$,
$\tau^{yy}_{j,0} := -\sigma^z_j$,
and $\tau^{xx}_{j,0} \equiv \tau^{yy}_{j,-1} := 0$.
Note that $Q_1^+ = H_1$.

Since the relations derived in Secs.~\ref{sec:PostQuenchTemperature} and~\ref{sec:ThermalizationConditions} do not depend explicitly on the system size,
we content ourselves with a relatively small value of $L = 10$ and use exact diagonalization for all numerical results in the following.

We first consider quenches in the external field.
The quench operator $V = M^z$ is the total $z$ magnetization,
\begin{equation}
\label{eq:Mz}
	M^z := \sum_j \sigma_j^z \,,
\end{equation}
and the pre-quench Hamiltonian $\tilde H = H - g V$ is thus of the same form~\eqref{eq:H:TFIM},
but with $h = \frac{1}{2} + g$.

To illustrate how Eq.~\eqref{eq:PostQuenchTemp:Expansion} 
can be used to predict the temperature after the quench,
we show in Fig.~\ref{fig:TFIM1}a the post-quench temperature $\beta$ against the pre-quench temperature $\tilde\beta$ for different strengths $g$
along with the zeroth-, first-, and second-order approximations [cf.\ Eqs.~\eqref{eq:PostQuenchTemp:0}, \eqref{eq:PostQuenchTemp:1}, and~\eqref{eq:PostQuenchTemp:2}].
As expected, the approximations work best for smaller quench strengths and higher temperatures (small $\beta$),
and the prediction improves as higher orders are included.
Qualitatively, we observe that the quench cools the system at high initial temperature (small $\tilde\beta$), i.e., $\beta \geq \tilde\beta$,
whereas it heats the system if starting from a low-temperature state (large $\tilde\beta$).
This trend is reflected correctly in the first- and second-order approximations.

Let us now turn to the behavior of the LIOM
and focus on $Q_n^+$ from~\eqref{eq:Q:TFIM:+} first.
In Fig.~\ref{fig:TFIM1}b,
we plot the expectation values $\< Q_n^+ \>_{\!\tilde\rho_{\tilde\beta}}$ for $n = 2, 3, 4$ in the pre-quench canonical ensemble as black solid lines
along with the first- and second-order approximations according to Eq.~\eqref{eq:PreQuench:Expv:Leading} as gray dashed and dot-dashed lines, respectively.
Since $\tilde\rho_{\tilde\beta}$ is the initial state and the $Q_n^+$ are conserved,
these are the actual expectation values assumed by the system after the quench.
We compare them to the expectation values $\< Q_n^+ \>_{\!\rho_\beta}$ of the post-quench canonical ensemble at the post-quench inverse temperature $\beta$ (which itself depends on $\tilde\beta$ as shown in Fig.~\ref{fig:TFIM1}a).
Those $\< Q_n^+ \>_{\!\rho_\beta}$ are depicted as blue solid lines in Fig.~\ref{fig:TFIM1}b
along with their first- and second-order approximations according to Eq.~\eqref{eq:PostQuench:Expv:Leading} as blue dashed and dot-dashed lines, respectively.
They represent the values expected for a system that thermalizes in the conventional sense,
meaning that its long-time properties are macroscopically indistinguishable from those of the canonical Gibbs ensemble.
Hence, the necessary thermalization condition~\eqref{eq:PostQuenchQ:Condition} is satisfied if the black and blue lines coincide.

The results in Fig.~\ref{fig:TFIM1}b demonstrate that
the first- and second-order approximations reproduce the true expectation values in the pre- and post-quench canonical ensembles as expected for sufficiently small $g$ and $\tilde\beta$,
and the same conclusion can be drawn from all further numerical examples below.

In the present setup, we observe that the thermalization conditions are generally violated for the $Q_n^+$.
In line with Ref.~\cite{Fagotti2013rdm},
	the violation decreases in absolute terms as $n$ increases,
	i.e., the ``more local'' conserved quantities with smaller $n$ are more strongly violated
	and thus more important when setting up a generalized Gibbs ensemble to capture the system's equilibrium properties.
The thermalization conditions are already violated at first order,
as can be seen by comparing the gray and blue dashed lines.
Hence the simple condition~\eqref{eq:CondTherm:1} [or~\eqref{eq:CondTherm:1:Geometric}]
is sufficient to conclude that the system will fail to thermalize to the canonical ensemble.
Geometrically,
this is an instance of the scenario sketched in Fig.~\ref{fig:LIOMCond:1:Geometric}a,
where none of the operators $H, Q, V, Q^\perp, V^\perp$ are pairwise orthogonal
with respect to the post-quench canonical covariance $\cov(\cdot, \cdot)$.

The situation is manifestly different for the second set of LIOM,
the $Q_n^-$ from~\eqref{eq:Q:TFIM:-}.
Contrary to the $Q_n^+$,
those are current-like observables;
e.g., $Q_1^- = -J^z$ is the nearest-neighbor-hopping current of the $z$ magnetization,
\begin{equation}
\label{eq:Jz}
	J^z := \frac{1}{2\I} \sum_j \left( \sigma_j^- \sigma_{j+1}^+ - \sigma_j^+ \sigma_{j+1}^- \right)
\end{equation}
with $\sigma_j^\pm := \sigma^x_j \pm \I \sigma^y_j$.

Denoting by $K$ the (antiunitary and self-inverse) operator defined by complex conjugation in the product basis,
we observe that $K H_1 K^{-1} = H_1$ and $K Q_n^- K^{-1} = -Q_n^-$,
from which we can conclude that $\tr[f(H_1) Q_n^-] = 0$ for an arbitrary function $f(x)$.
(The latter relation can also be understood straightforwardly from the ``diagonal'' representations of $H_1$ and $Q_n^-$ in terms of Bogoliubov operators \cite{Fagotti2013rdm}.)
It follows that $\< Q_n^- \>_{\!\tilde\rho_{\tilde\beta}} = \< Q_n^- \>_{\!\rho_\beta} = 0$
(hence they are not shown explicitly in Fig.~\ref{fig:TFIM1}).
Consequently,
the condition~\eqref{eq:PostQuenchQ:Condition} is satisfied (at all orders in $g$)
and the $Q_n^-$ are irrelevant in this scenario.
At first order, the geometric interpretation is a special case of Fig.~\ref{fig:LIOMCond:1:Geometric}c with $Q = Q^\perp$.

\begin{figure}
	\includegraphics[scale=1]{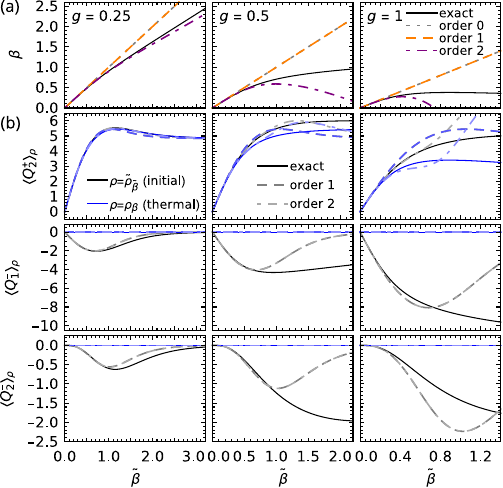}
	\caption{(a) Post-quench inverse temperature $\beta$ and (b) LIOM expectation values $\<Q^\pm_n\>_{\!\rho}$ as a function of the pre-quench inverse temperature $\tilde\beta$
		for canonical quenches to the transverse-field Ising model $H = H_1$ from~\eqref{eq:H:TFIM} with $h = \frac{1}{2}$.
		Pre-quench: $\tilde H = H_1 - g J^z$ with $g = 0.25$ (left), $0.5$ (middle), $1$ (right).
		Line styles and colors as indicated in the legends and the same as in Fig.~\ref{fig:TFIM1}.}
	\label{fig:TFIM2}
\end{figure}

We contrast this with a second scenario
where the quench operator $V = J^z$ is the spin current from~\eqref{eq:Jz}.
Temperatures and LIOM expectation values for various quench strengths are shown in Fig.~\ref{fig:TFIM2}.
In contrast to the previous case, the quench now always heats up the system ($\beta \leq \tilde\beta$).
Furthermore, the first-order correction~\eqref{eq:PostQuenchTemp:1} of the inverse temperature vanishes because $\cov(H, V) = 0$,
such that the gray dotted and orange dashed lines in Fig.~\ref{fig:TFIM2}a agree.
However, the second-order term~\eqref{eq:PostQuenchTemp:2} yields a finite correction.

The $Q_n^-$ now generally violate the first-order thermalization condition~\eqref{eq:CondTherm:1}
as illustrated for $Q_1^-$ and $Q_2^-$ in Fig.~\ref{fig:TFIM2}b \cite{note:Q1-Quench}.
Geometrically, this can be regarded as a special case of the situation in Fig.~\ref{fig:LIOMCond:1:Geometric}a with $Q = Q^\perp$ and $V = V^\perp$ (and $Q = -V$ for $Q_1^-$).
Interestingly, the second-order corrections to the expectation values in~\eqref{eq:PostQuench:Expv:Leading} and~\eqref{eq:PreQuench:Expv:Leading}
vanish (such that the dashed and dot-dashed lines coincide),
hence the second-order condition~\eqref{eq:CondTherm:2} is fulfilled.
By contrast (and as exemplified for $Q_2^+$),
the $Q_n^+$ all satisfy the first-order condition~\eqref{eq:CondTherm:1}
in this setup because $\cov(Q_n^+, V) = \cov(H, V) = 0$
(see above; special case of Fig.~\ref{fig:LIOMCond:1:Geometric}c with $V = V^\perp$),
but violate the second-order condition~\eqref{eq:CondTherm:2}.
Overall, all LIOM of the transverse-field Ising model are thus relevant in this scenario,
and even though this cannot be detected at first order alone,
it becomes apparent at second order in the quench strength.

\subsection{Nonintegrable XXZ model}

\begin{figure}
	\includegraphics[scale=1]{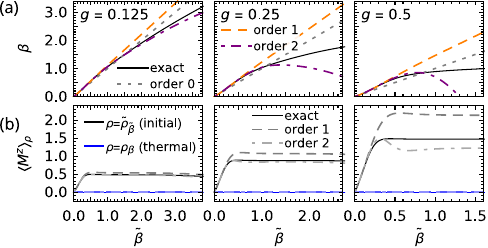}
	\caption{(a) Post-quench inverse temperature $\beta$ and (b) $z$-magnetization expectation values $\<M^z\>_{\!\rho}$
		as a function of the pre-quench inverse temperature $\tilde\beta$
		for canonical quenches in the nonintegrable XXZ model.
		Post-quench Hamiltonian $H = H_2$ from~\eqref{eq:H:XXZ+},
		quench operator $V = W$ from~\eqref{eq:V:XXZ+:9S},
		and strengths $g = 0.125$ (left), $0.25$ (middle), $0.5$ (right).
		Line styles and colors as indicated in the legends and the same as in Fig.~\ref{fig:TFIM1}.
	}
	\label{fig:XXZ+9S}
\end{figure}

\begin{figure}
	\includegraphics[scale=1]{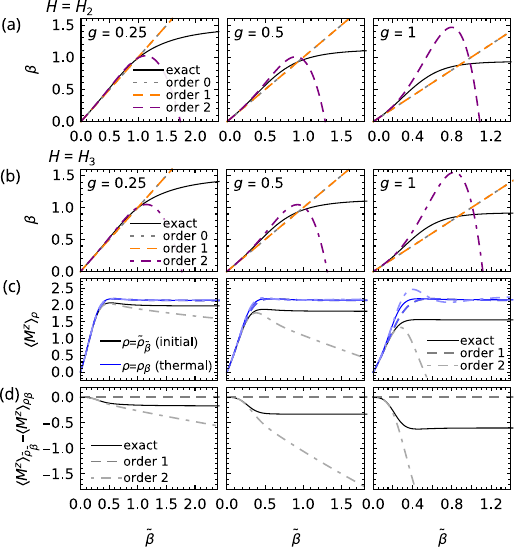}
	\caption{(a) Post-quench inverse temperature $\beta$
		as a function of the pre-quench inverse temperature $\tilde\beta$
		for canonical quenches in the nonintegrable XXZ model with post-quench Hamiltonian $H = H_2$ from~\eqref{eq:H:XXZ+},
		quench operator $V = M^x$ from~\eqref{eq:V:XXZ+:5},
		and quench strengths $g = 0.25$ (left), $0.5$ (middle), $1$ (right).
		(b) Same relationship between pre- and post-quench temperatures,
		but for the post-quench Hamiltonian $H = H_3$ from~\eqref{eq:H:XXZ+:5Sa}.
		(c) $z$-magnetization expectation values $\< M^z \>_{\!\rho}$ for the initial ($\tilde\rho_{\!\tilde\beta}$) and thermal ($\rho_\beta$) states,
		and (d) their difference,
		for the scenario with post-quench Hamiltonian $H = H_3$.
		Line styles and colors as indicated in the legends and the same as in Fig.~\ref{fig:TFIM1}.
	}
	\label{fig:XXZ+5S}
\end{figure}

Our second set of examples
involves a variant of the XXZ model with nearest- and next-nearest-neighbor couplings.
For the post-quench system,
we first choose $H = H_2$, where
\begin{equation}
\label{eq:H:XXZ+}
	H_2 := -\sum_{j=1}^L \sum_{n=1}^2 \left(
		\sigma_j^x \sigma_{j+n}^x
		+ \sigma_j^y \sigma_{j+n}^y
		+ \Delta \sigma_j^z \sigma_{j+n}^z
	\right)
\end{equation}
with $\Delta = \frac{1}{2}$.
This model is provably nonintegrable in the sense that the number of local conserved quantities is finite in the thermodynamic limit \cite{Shiraishi2024alc}.
The only (known) local conserved quantity besides $H$ is the total $z$ magnetization $Q = M^z$ from~\eqref{eq:Mz}.

First,
we inspect quenches from a system with Hamiltonian $\tilde H = H_2 - g W$
where
\begin{equation}
	\label{eq:V:XXZ+:9S}
	W := \sum_j \left( \sigma_j^x \sigma_{j+3}^x + \sigma_j^y \sigma_{j+3}^y + \sigma_j^z \right) ,
\end{equation}
i.e., a model with additional third-nearest neighbor $x$ and $y$ couplings and a magnetic field applied in the $z$ direction.

In this case, $H$ and $Q$ are orthogonal to each other [$\cov(H_2, M^z) = 0$],
but $V = W$ is not orthogonal to either of $H$ or $Q$.
This results in finite corrections to the post-quench temperature and violations of the thermalization conditions at both first and second order,
visualized in Fig.~\ref{fig:XXZ+9S}.
Consequently,
a 
generalized Gibbs ensemble that takes into account the conserved $z$ magnetization is necessary to describe the system's equilibrium properties correctly.

Next,
we prepare the system before the quench in an equilibrium state with an additional magnetic field in the $x$ direction instead.
The quench operator $V = M^x$ is thus the $x$ magnetization
\begin{equation}
\label{eq:V:XXZ+:5}
	M^x := \sum_j \sigma_j^x \,.
\end{equation}
Since both $H = H_2$ and $V = M^x$ are spin-flip symmetric in the $z$ direction
(they commute with $F^z = \prod_{j=1}^L \sigma_j^x$),
we have $\tr[f(H, V) Q] = 0$ for an arbitrary function $f(x, y)$,
hence $\cov(H, Q) = \cov(Q, V) = 0$.
Furthermore, $H$ is also spin-flip symmetric in the $x$ direction (commutes with $F^x = \prod_{j=1}^L \sigma_j^z$),
such that $\cov(H, V) = 0$.
In conclusion,
$H$, $V$, and $Q$ are all mutually orthogonal to each other with respect to the canonical-ensemble covariance;
cf.\ Fig.~\ref{fig:LIOMCond:1:Geometric}c with $V = V^\perp$ and $Q = Q^\perp$.
Hence, the first-order correction of the post-quench temperature vanishes
as shown in Fig.~\ref{fig:XXZ+5S}a (but the second order provides a nontrivial correction).

On the other hand,
the spin-flip symmetry also implies that
the $z$ magnetization vanishes in both the pre- and post-quench canonical ensembles,
$\< M^z \>_{\!\tilde\rho_{\tilde\beta}} = \< M^z \>_{\!\rho_\beta} = 0$,
such that the thermalization condition~\eqref{eq:PostQuenchQ:Condition}
is satisfied to all orders.
The canonical ensemble without reference to the conserved $z$ magnetization thus reproduces the equilibrium properties of the post-quench system.

Interestingly,
this changes if we add the same, constant magnetic field to both the pre- and post-quench Hamiltonians,
i.e.,
we consider $H = H_3$ with
\begin{equation}
\label{eq:H:XXZ+:5Sa}
	H_3 := H_2 - h M^z
\end{equation}
and the same quench operator $V = M^x$.
Since $H$ still commutes with $F^x$,
$\cov(H, V) = 0$ continues to hold.
Likewise, we still have $\cov(Q, V) = 0$,
but $H$ and $Q$ are no longer orthogonal in general.
This means that the first-order temperature correction still vanishes,
and in fact, the relationship between the pre- and post-quench temperatures is essentially unaffected by the additional magnetic field
as can be seen by comparing Figs.~\ref{fig:XXZ+5S}a and b.
The expectation values of the conserved $z$ magnetization, however,
are affected and now generally differ in the pre- and post-quench ensembles
even though the additional field $h$ is the same in both $H = H_3 = H_2 - h M^z$ and $\tilde H = H_3 - g M^x = H_2 - h M^z - g M^x$.
Despite the quench operator being orthogonal to both $H$ and $M^z$,
the nontrivial equilibrium $z$ magnetization induced by the field $h$ depends on the field strength $g$ in the $x$ direction.
The corresponding expectation values of $M^z$ are shown in Fig.~\ref{fig:XXZ+5S}c
along with their difference $\< M^z \>_{\!\tilde\rho_{\tilde\beta}} - \< M^z \>_{\!\rho_\beta}$ in Fig.~\ref{fig:XXZ+5S}d.
The first-order thermalization condition~\eqref{eq:CondTherm:1} fails to capture this qualitative change between the $H = H_2$ and $H = H_3$ scenarios
(the dashed curve in Fig.~\ref{fig:XXZ+5S}d indicates zero violation for all $\tilde\beta$),
but the second-order condition~\eqref{eq:CondTherm:2} indicates a violation
(the dot-dashed line in Fig.~\ref{fig:XXZ+5S}d shows a nonzero deviation).

\section{Discussion and conclusions}
\label{sec:Conclusions}

Thanks to their experimental practicability,
quenches from finite-temperature thermal equilibrium states have been of significant interest to understand the relaxation behavior of many-body systems theoretically (see, e.g., Refs.~\cite{Rigol2014qqt, Rigol2016faq, Farrelly2017tre, Mori2017tet, Mallayya2018qqr, Richter2019iet, Dabelow2022tlp, Varizi2022qqt} and references therein). 
Considering such quenches of a quantum system from a canonical equilibrium state,
we derived systematic expansions in the quench strength to relate equilibrium properties of the post-quench system to equilibrium properties at its pre-quench temperature.
The key physical insights are
(i) an estimate of the post-quench temperature [Eqs.~\eqref{eq:PostQuenchTemp:Expansion}, \eqref{eq:PostQuenchTemp:1}, \eqref{eq:PostQuenchTemp:2}, and~\eqref{eq:PostQuenchTemp:n}];
(ii) a prediction of post-quench equilibrium observable expectation values in thermalizing systems [Eqs.~\eqref{eq:PostQuench:Expv:Leading} and~\eqref{eq:PostQuench:Expv2}];
(iii) a hierarchy of necessary conditions for the system to thermalize to the canonical ensemble after the quench in the presence of additional conserved quantities [Eqs.~\eqref{eq:CondTherm:1}, \eqref{eq:CondTherm:2}, \eqref{eq:PostQuench:Expv2}, \eqref{eq:PreQuench:Expv}].

At sufficiently high temperatures and small quench strengths,
low-order truncations of the expansions~(i) and~(ii) offer a simple and reliable way to approximate the post-quench temperature and, for a thermalizing system, expected late-time values of observables.
This was demonstrated numerically for various scenarios in Figs.~\ref{fig:TFIM1}--\ref{fig:XXZ+5S} (see also the summary in Tab.~\ref{tab:Examples}).
Note that the opposite low-temperature regime and specifically quenches from the ground state of a pre-quench Hamiltonian,
another frequently studied setup \cite{Essler2016qdr, Mitra2018qqd},
are not directly accessible quantitatively (unless one goes to very high orders in the expansion or very small quench amplitudes),
but qualitative results like the thermalization conditions~(iii) remain useful as they must be satisfied at all orders.

The thermalization conditions~(iii) were obtained by
comparing the pre- and post-quench expectation values of (physically observable) conserved quantities at each order in the quench strength. They
can be seen as kinematic constraints
and can be tested solely in terms of equilibrium properties,
namely canonical-ensemble averages of the post-quench system at the pre-quench temperature. 
They do not depend on or require access to any dynamical properties such as time-dependent or time-averaged expectation values.
Likewise, they do not involve concepts like quantum chaos,  ergodicity, or eigenstate thermalization \cite{Deutsch1991qsm, Srednicki1994cqt, Rigol2008tim}.
Then again, they are only necessary conditions,
suitable for detecting the absence of thermalization,
but not for guaranteeing its presence. 

An interesting complementary approach by Chiba and Shimizu \cite{Chiba2023kol}
identified a sufficient condition for thermalization at first order in the quench strength.
In particular,
they showed that time-averaged expectation values of arbitrary (sums of) local observables coincide with their thermal, canonical expectation values to first order in $g$
if and only they do so for the quench operator $V$ in our context.
For a conserved quantity,
this agreement between time-averaged and thermal expectation values boils down to
Eq.~\eqref{eq:PostQuenchQ:Condition}.

Importantly, however, we observed (see Fig.~\ref{fig:XXZ+5S}b-d) that the system may appear to thermalize at first order,
in the sense that the necessary condition~\eqref{eq:CondTherm:1} holds,
but is revealed to violate higher-order conditions.

Another main physical insight follows from the observation that
the first-order thermalization condition
is explicitly violated if the quench operator is a conserved quantity of the post-quench
system:
This offers
an operational answer to the question of what constitutes a potentially relevant conserved quantity,
namely any observable that commutes with the Hamiltonian and is a legitimate quench operator.
In other words,
any conserved quantity for which there is an experimentally accessible conjugate control parameter will be relevant (at least) in an associated experiment where that parameter is quenched.

An otherwise commonly adopted characterization of relevant conserved quantities in theory
is (sums of) local or quasilocal operators \cite{Gogolin2016ete, DAlessio2016qce, Essler2016qdr}.
However, there may be other types of conserved quantities (e.g., momentum) that are controllable and potentially relevant \cite{Kinoshita2006qnc, Gangardt2008ceg, Santos2011eiq}.
A closely related question is what constitutes a ``physical observable,''
such that one generically expects its long-time expectation values to approach the values predicted by a pertinent thermodynamic ensemble.
How this class is chosen clearly affects what kinds of conserved quantities need to be admitted when defining an equilibrium ensemble.

Finally,
we remark that
the approach adopted here can be employed more generally to expand a 
generalized Gibbs ensemble in all Lagrange parameters (generalized inverse temperatures).
The contributions to these parameters at every order in the quench strength can then be calculated recursively by solving a linear system whose coefficients depend on lower-order parameters as well as expectation values and correlation functions of a post-quench ensemble at the pre-quench parameter values.
Furthermore, given the commonly expected and often provable equivalence of thermodynamic ensembles \cite{Touchette2015ene, Brandao2015esm, Tasaki:2018lec, Kuwahara2020etc, Kuwahara2020gcb},
it is expected that similar thermalization conditions can be formulated in the microcanonical picture,
where relevant conserved quantities are constrained by restricting the Hilbert space instead of fixing mean values through Lagrange multipliers.
In case of a finite number of additional discrete and mutually commuting local conserved quantities,
one may alternatively break down the analysis into symmetry subsectors
and use the ``one-parameter'' canonical ensemble as adopted in this work
in every subsector.

\begin{acknowledgments}
I am grateful to Peter Reimann for numerous inspiring discussions on this and related topics.
\end{acknowledgments}


\appendix

\section{Canonical ensembles at different temperatures}
\label{app:DifferentTemperatures}

In this section,
we derive a perturbative expansion that relates canonical expectation values of an observable $A$ at two different inverse temperatures $\beta$ and $\beta_0$.
We think of $\beta_0$ as the pre-quench and $\beta$ as the post-quench temperature.

Our starting point is the expansion~\eqref{eq:PostQuenchTemp:Expansion} for $\beta - \beta_0 = \sum_{n=1}^\infty \beta_n g^n$,
which we substitute into the canonical exponential,
\begin{equation}
\label{eq:PostQuench:Exp:0}
	\e^{-(\beta - \beta_0) H}
		= \sum_{n=0}^\infty \frac{(-H)^n}{n!} \left[ \sum_{k=1}^\infty \beta_k g^k \right]^n \,.
\end{equation}
Expanding the multinomial, we obtain
\begin{equation}
\label{eq:PostQuench:Exp:1}
	\e^{-(\beta - \beta_0) H}
		= \sum_{n=0}^\infty \frac{(-H)^n}{n!} \sum_{\bm\alpha : \lvert\bm\alpha\rvert = n} \frac{n!}{\bm\alpha!} \prod_{k=1}^\infty (\beta_k g^k)^{\alpha_k} \,,
\end{equation}
where the inner sum is over all multiindices $\bm\alpha = (\alpha_1, \alpha_2, \ldots)$ with $\alpha_k \in \NN_0$ (nonnegative integers) and length $\lvert\bm\alpha\rvert := \sum_{k=1}^\infty \alpha_k = n$.
The factorial of a multiindex is defined as $\bm\alpha! := \prod_{k=1}^\infty \alpha_k!$.
Since the double sum in~\eqref{eq:PostQuench:Exp:1} effectively generates all multiindices,
we can write the relation more compactly as
\begin{equation}
\label{eq:PostQuench:Exp:2}
	\e^{-(\beta - \beta_0) H} = \sum_{\bm\alpha} \frac{(-H)^{\lvert\bm\alpha\rvert}}{\bm\alpha!} \prod_{k=1}^\infty (\beta_k g^k)^{\alpha_k} \,.
\end{equation}
Next, we introduce the set of integer partitions
\begin{equation}
\label{eq:IntegerPartition}
	\Part(n) := \left\{ (\alpha_1, \alpha_2, \ldots) \in \NN_0^\infty : \sum_{k=1}^n k \, \alpha_k = n \right\}
\end{equation}
with $\Part(0) := \{ \emptyset \}$ by convention (the one-element set containing only the empty set).
Note that, given $\bm\alpha \in \Part(n)$, $\alpha_k$  is necessarily zero whenever $k > n$;
hence, we also write the elements of $\Part(n)$ as finite $n$-tuples.
These elements of $\Part(n)$ encode all possible ways of writing $n$ as a sum of positive integers.
For example, $\Part(3)$ consists of $(3, 0, 0)$, $(1, 1, 0)$, and $(0, 0, 1)$, representing the partitions $1 + 1 + 1$, $1 + 2$, and $3$, respectively.

We can now decompose the sum over all multiindices in~\eqref{eq:PostQuench:Exp:2} into integer partitions,
which effectively sorts the exponential by powers of $g$,
\begin{equation}
\label{eq:PostQuench:Exp:Final}
	\e^{-(\beta - \beta_0) H} = \sum_{n=0}^\infty g^n \sum_{\bm\alpha \in \Part(n)} \frac{(-\bm\beta)^{\bm\alpha}}{\bm\alpha!} H^{\lvert\bm\alpha\rvert}
\end{equation}
with $\bm\beta := (\beta_1, \beta_2, \ldots)$ and the short-hand notation
\begin{equation}
\label{eq:MultiindexPowers}
	\bm\beta^{\bm\alpha} := \prod_{k=1}^\infty \beta_k^{\alpha_k} \,.
\end{equation}

Turning to the partition function $Z_\beta$ (cf.\ Eq.~\eqref{eq:rho0}),
we multiply~\eqref{eq:PostQuench:Exp:Final} by $\e^{-\beta_0 H}$ and take the trace,
\begin{equation}
	Z_\beta = \sum_{n=0}^\infty g^n \sum_{\bm\alpha \in \Part(n)} \frac{(-\bm\beta)^{\bm\alpha}}{\bm\alpha!} \tr(\e^{-\beta_0 H} H^{\lvert\bm\alpha\rvert}) \,.
\end{equation}
Dividing by $Z_{\beta_0}$,
we can rewrite this in terms of expectation values of the post-quench ensemble at the pre-quench (or zeroth-order post-quench) temperature,
\begin{equation}
\label{eq:PostQuench:PartitionFunction}
	\frac{Z_\beta}{Z_{\beta_0}}
		 = 1 + \sum_{n=1}^\infty g^n z_n(\bm\beta)
\end{equation}
with
\begin{equation}
	z_n(\bm\beta) := \sum_{\bm\alpha \in \Part(n)} \frac{(-\bm\beta)^{\bm\alpha}}{\bm\alpha!} \< H^{\lvert\bm\alpha\rvert} \>_{\rho_{\beta_0}}
\end{equation}
To keep the notation compact,
we will usually suppress the dependence of $z_n$ on $\bm\beta$ in the following.

To calculate expectation values in the canonical ensemble,
we need the reciprocal of~\eqref{eq:PostQuench:PartitionFunction}.
Using the geometric series expansion and following the same steps that led from~\eqref{eq:PostQuench:Exp:0} to~\eqref{eq:PostQuench:Exp:Final},
we obtain
\begin{equation}
\label{eq:PostQuench:PartitionFunction:Reciprocal}
	\frac{Z_{\beta_0}}{Z_\beta}
		= \sum_{n=0}^\infty (-1)^n \left[ \sum_{k=1}^\infty g^k z_k \right]^n
		= \sum_{n=0}^\infty g^n \!\! \sum_{\bm\alpha \in \Part(n)} \!\! \frac{\lvert\bm\alpha\rvert! \, (-\bm z)^{\bm\alpha}}{\bm\alpha!} \,,
\end{equation}
where $\bm z := (z_1, z_2, \ldots)$.
Combining Eqs.~\eqref{eq:PostQuench:Exp:Final} and~\eqref{eq:PostQuench:PartitionFunction:Reciprocal},
we find that the canonical expectation value of an observable $A$ can be written as
\begin{equation}
\label{eq:PostQuench:Expv1}
\begin{aligned}
	\< A \>_{\!\rho_\beta}
		&= \left[ \sum_{m=0}^\infty g^m \!\! \sum_{\bm\mu \in \Part(m)} \!\! \frac{\lvert\bm\mu\rvert! \, (-\bm z)^{\bm\mu}}{\bm\mu!} \right]
			\\ & \quad \times
			\left[ \sum_{n=0}^\infty g^n \!\! \sum_{\bm\nu \in \Part(n)} \!\! \frac{ (-\bm\beta)^{\bm\nu} }{ \bm\nu! } \< H^{\lvert\bm\nu\rvert} A \>_{\!\rho_{\beta_0}} \right] .
\end{aligned}
\end{equation}
Collecting powers of $g$,
we arrive at
\begin{equation}
\label{eq:PostQuench:Expv2}
	\< A \>_{\!\rho_\beta}
	=
	\sum_{n=0}^\infty g^n \sum_{k=0}^n \sum_{\bm\mu \in \Part(n-k)} \!\!\!\!\!\!\! \frac{\lvert\bm\mu\rvert! \, (-\bm z)^{\bm\mu}}{\bm\mu!} \!\!\! \sum_{\bm\nu \in \Part(k)} \!\!\! \frac{ (-\bm\beta)^{\bm\nu} }{ \bm\nu! } \< H^{\lvert\bm\nu\rvert} A \>_{\!\rho_{\beta_0}}
\end{equation}
Hence, we have expressed expectation values at inverse temperature $\beta$ in terms of expectation values at inverse temperature $\beta_0$,
where the two temperatures are related through the perturbative expansion~\eqref{eq:PostQuenchTemp:Expansion}.

\section{Canonical ensembles of different Hamiltonians}
\label{app:DifferentHamiltonians}

In this section,
we express the canonical expectation values of an observable $A$ with respect to a Hamiltonian $\tilde H$ in terms of canonical expectation values of a different Hamiltonian $H$.
We think of $\tilde H$ as the pre-quench Hamiltonian and $H = \tilde H + g V$ as the post-quench Hamiltonian.

The idea is to expand the exponential $\e^{-\tilde\beta \tilde H} = \e^{-\tilde\beta (H - g V)}$ in the quench strength $g$.
To this end, we define the operator
\begin{equation}
\label{eq:Phi}
	\Phi(\tilde\beta) := \e^{\tilde\beta H} \e^{-\tilde\beta (H - g V)}
\end{equation}
and observe that it satisfies the integral relation
\begin{equation}
\label{eq:Phi:Integral}
	\Phi(\tilde\beta) = 1 + g \int_0^{\tilde\beta} \d\lambda \, \e^{\lambda H} V \e^{-\lambda H} \Phi(\lambda) \,,
\end{equation}
as can be verified straightforwardly by differentiating with respect to $\tilde\beta$ and observing $\Phi(0) = 1$ \cite{Reimann2024orh}.
With the definitions
\begin{align}
	\Phi_0(\tilde\beta) &:= 1 \,,\\
	\Phi_n(\tilde\beta) &:= \int_0^{\tilde\beta} \d\lambda \, V(\lambda) \Phi_{n-1}(\lambda) \\
		&\;= \int_0^{\tilde\beta} \!\! \d\lambda_1 \int_0^{\lambda_1} \!\! \d\lambda_2 \cdots \int_0^{\lambda_{n-1}} \!\! \d\lambda_n \, V(\lambda_1) \cdots V(\lambda_n) \,,\\
	V(\lambda) &:= \e^{\lambda H} V \e^{-\lambda H} \,,
\end{align}
we thus obtain a perturbative expansion of $\Phi(\tilde\beta) = \sum_{n=0}^\infty \Phi_n(\tilde\beta) g^n$ in $g$.

The partition function $\tilde Z_{\tilde\beta} := \tr(\e^{-\tilde\beta \tilde H})$ of the pre-quench Hamiltonian $\tilde H$ can thus be written as
\begin{equation}
	\tilde Z_{\tilde\beta} = \tr e^{-\tilde\beta H} + \sum_{n=1}^\infty g^n \tr[ \e^{-\tilde\beta H} \Phi_n(\tilde\beta) ] \,.
\end{equation}
Identifying $Z_{\tilde\beta} := \tr \e^{-\tilde\beta H}$ as the partition function of the post-quench Hamiltonian $H$ at inverse temperature $\tilde\beta$,
we can then express $\tilde Z_{\tilde\beta}$ in terms of correlation functions in the post-quench canonical ensemble,
\begin{equation}
	\frac{\tilde Z_{\tilde\beta}}{Z_{\tilde\beta}}
		= 1 + \sum_{n=1}^\infty g^n \< \Phi_n(\tilde\beta) \>_{\! \rho_{\tilde\beta}} \,.
\end{equation}
To obtain the reciprocal of this relation,
we apply the geometric series expansion once again (cf.\ above Eq.~\eqref{eq:PostQuench:PartitionFunction:Reciprocal})
and find
\begin{equation}
\label{eq:PreQuench:PartitionFunction:Reciprocal}
	\frac{Z_{\tilde\beta}}{\tilde Z_{\tilde\beta}}
		= \sum_{n=0}^\infty g^n \sum_{\bm\alpha \in \Part(n)} \frac{\lvert \bm\alpha \rvert !}{\bm\alpha !} \prod_{k=1}^n \< -\Phi_k(\tilde\beta) \>_{\! \rho_{\tilde\beta}}^{\,\alpha_k} \,.
\end{equation}
To calculate expectation values in the pre-quench canonical ensemble
in terms of post-quench expectation values and correlation functions,
we observe that
\begin{equation}
\label{eq:PreQuench:ExpvNum}
	\tr(\e^{-\tilde\beta \tilde H} A) = \tr[ \e^{-\tilde\beta H} \Phi(\tilde\beta) A ]
		= \sum_{n=0}^\infty g^n \tr[ \e^{-\tilde\beta H} \Phi_n(\tilde\beta) A ] \,.
\end{equation}
Combining Eqs.~\eqref{eq:PreQuench:PartitionFunction:Reciprocal} and~\eqref{eq:PreQuench:ExpvNum}
and collecting powers of $g$ (cf.\ Eqs.~\eqref{eq:PostQuench:Expv1} and~\eqref{eq:PostQuench:Expv2}),
we obtain
\begin{equation}
\label{eq:PreQuench:Expv}
	\< A \>_{\! \tilde\rho_{\tilde\beta}}
		= \sum_{n=0}^\infty g^n \sum_{k=0}^n \< \Phi_k(\tilde\beta) A \>_{\!\rho_{\tilde\beta}}\!\!  \sum_{\mu \in \Part(n-k)} \!\! \frac{\lvert \bm\mu \rvert!}{\bm\mu!} \prod_{j=1}^{n-k} \< -\Phi_j(\tilde\beta) \>_{\!\rho_{\tilde\beta}}^{\,\mu_j} \,.
\end{equation}
Thus, we have expanded pre-quench canonical expectation values into a perturbation series in the quench strength $g$,
which involves post-quench canonical expectation values and imaginary-time correlation functions.

\section{Post-quench temperature}
\label{app:PostQuenchTemperature}

If the system is prepared in the canonical ensemble $\tilde\rho_{\tilde\beta}$ of the pre-quench Hamiltonian $\tilde H$ at inverse temperature $\tilde\beta$,
its inverse temperature $\beta$ after a quench to the Hamiltonian $H$
is determined by the condition~\eqref{eq:PostQuenchTemp:Condition}, $\<H\>_{\!\tilde\rho_{\tilde\beta}} = \<H\>_{\!\rho_\beta}$.
In other words, the energy  $\<H\>_{\!\tilde\rho_{\tilde\beta}}$ of the initial state has to equal the energy $\<H\>_{\!\rho_\beta}$ in the post-quench canonical ensemble.
Using Eqs.~\eqref{eq:PostQuench:Expv2} and~\eqref{eq:PreQuench:Expv},
we can thus calculate the post-quench temperature order by order in the quench amplitude $g$.

At zeroth order ($g = 0$), we find that $\<H\>_{\!\tilde\rho_{\tilde\beta}} = \< H \>_{\!\rho_{\tilde\beta}}$ and $\< H \>_{\!\rho_\beta} = \< H \>_{\!\rho_{\beta_0}}$.
Consequently,
\begin{equation}
	\beta_0 = \tilde\beta \,,
\end{equation}
i.e., the zeroth-order approximation of the post-quench temperature is the pre-quench temperature as expected.

In the following, we will use the shorthand notation
introduced below Eq.~\eqref{eq:PostQuench:Expv:Leading},
$\< A \> := \< A \>_{\!\rho_{\beta_0}} \equiv \< A \>_{\!\rho_{\tilde\beta}}$ for post-quench canonical expectation values at the pre-quench temperature
as well as $\cov(A, B) := \< A B \> - \< A \> \< B \>$ and $\var(A) := \cov(A, A)$.
Furthermore, we define $\bm\Phi := (\Phi_1(\beta_0), \Phi_2(\beta_0), \ldots)$ and recall the notation from~\eqref{eq:MultiindexPowers} for ``sequences to the power of multiindices.''

Comparing Eqs.~\eqref{eq:PostQuench:Expv2} and~\eqref{eq:PreQuench:Expv} at order $n$,
we find that $\beta_n$ can be calculated recursively as follows:
\begin{widetext}
\begin{equation}
\label{eq:PostQuenchTemp:n}
\begin{aligned}
	\beta_n \var(H)
		&= 
		-\cov(\Phi_n(\beta_0), H)
		+ \!\! \sum_{\bm\mu \in \Part'(n)} \! \frac{1}{\bm\mu!} \left\{ \lvert\bm\mu\rvert! \< H \> \left[ (-\bm z)^{\bm\mu} - \< -\bm\Phi \>^{\bm\mu} \right] + \cov(H^{\lvert\mu\rvert}, H) (-\bm\beta)^{\bm\mu} \right\} \\
		&\qquad + \sum_{k=1}^{n-1} \sum_{\bm\mu \in \Part(n-k)} \!\!\!\! \frac{ \lvert\bm\mu\rvert! (-\bm z)^{\bm\mu} }{ \bm\mu! } \!\! \sum_{\bm\nu \in \Part(k)} \!\! \frac{ (-\bm\beta)^{\bm\nu} \< H^{\lvert\bm\nu\rvert + 1} \> }{ \bm\nu! } \,,
\end{aligned}
\end{equation}
\end{widetext}
where $\Part'(n)$ denotes the set of all integer partitions of $n$ except $(0, \ldots, 0, 1)$ ($\equiv n$).
The first-order correction~\eqref{eq:PostQuenchTemp:1} follows by observing that the sums on the right-hand side of Eq.~\eqref{eq:PostQuenchTemp:n} are empty if $n = 1$ and
\begin{equation}
	\cov(\Phi_1(\beta_0), H)
		= \beta_0 \cov(H, V) \,.
\end{equation}
The second-order correction~\eqref{eq:PostQuenchTemp:2}
is obtained by using $\mathcal{P}(0) = \{\emptyset\}$, $\mathcal{P}(1) = \{ (1) \}$,
$\Part(2) = \{(2, 0), (0, 1)\}$, and $\Part'(2) = \{(2, 0)\}$.

\section{Properties of the canonical covariance}
\label{app:CovarianceInnerProduct}

For completeness,
we collect a few properties of the canonical covariance
\begin{equation}
	\cov(A, B) := \tr(\rho_{\beta_0} A B) - \tr(\rho_{\beta_0} A) \tr(\rho_{\beta_0} B)
\end{equation}
that establish its similarity to an inner product on $\mathcal{L}_H(\mathcal{H})$,
the vector space of self-adjoint operators on the underlying Hilbert space $\mathcal{H}$.

\emph{Bilinearity.}
Let $A, A_1, A_2, B, B_1, B_2 \in \mathcal{L}_H(\mathcal{H})$ and $\lambda_1, \lambda_2 \in \RR$.
Then
\begin{equation}
	\cov(\lambda_1 A_1 + \lambda_2 A_2, B)
		= \lambda_1 \cov(A_1, B) + \lambda_2 \cov(A_2, B)
\end{equation}
and
\begin{equation}
	\cov(A, \lambda_1 B_1 + \lambda_2 B_2) = \lambda_1 \cov(A, B_1) + \lambda_2 \cov(A, B_2) \,.
\end{equation}

\emph{Positive definiteness.}
Let $A \in \mathcal{L}_H(\mathcal{H})$.
We observe that
\begin{align}
\label{eq:CanonicalCovariance:PositiveDefiniteness}
	\cov(A, A) 
		&= Z_{\beta_0}^{-1} \tr[ \e^{-\beta_0 H} (A - \< A \>)^2 ] \notag \\
		&= Z_{\beta_0}^{-1} \lVert \e^{-\beta_0 H/2} (A - \<A\>) \rVert^2_{\mathrm{F}} \,,
\end{align}
where $\lVert X \rVert_{\mathrm{F}} := \sqrt{\tr(X^\dagger X)}$ denotes the Frobenius norm of the (not necessarily self-adjoint) operator $X$.
Hence $\cov(A, A) \geq 0$,
which establishes that the canonical covariance is positive semidefinite.
Furthermore,
Eq.~\eqref{eq:CanonicalCovariance:PositiveDefiniteness} implies that $\cov(A, A) = 0$ if and only if $\e^{-\beta_0 H/2} (A - \<A\>) = 0$.
Since $\e^{-\beta_0 H/2}$ is a positive operator,
the latter condition is equivalent to $A = \< A \>$.
Hence the canonical covariance is positive definite on $\mathcal{L}_H(\mathcal{H}) / \RR$,
i.e., on the quotient space of self-adjoint operators modulo the real numbers.


%


\begin{thebibliography}{68}
\bibitem [{not({\natexlab{a}})}]{note:CanonicalEnsemble}%
	\BibitemOpen
	\href@noop {} {}
	\bibinfo {note} {At this point, we do not
		make any assumptions about the existence of additional conserved quantities
		besides $H$. In particular, the concept of a ``particle number'', which is
		typically associated with textbook discussions of the canonical ensemble, may
		or may not exist, and if it exists, it may or may not be conserved. In other words, we will
		use the term ``canonical ensemble'' somewhat loosely for states that are
		defined precisely by Eq.~\eqref{eq:rho0}.}\BibitemShut {Stop}%
\bibitem [{not({\natexlab{b}})}]{note:Q1-Quench}%
	\BibitemOpen
	\href@noop {} {} 
	\bibinfo {note} {Recall that $Q_n^{-1}$
		is analytically known to violate the first-order condition in this scenario
		because it coincides with the quench operator (up to a minus sign); see the
		discussion around Eq.~\eqref{eq:CondTherm:1:Q=V}.}\BibitemShut {Stop}%
\end{thebibliography}
\end{document}